\documentclass[journal=jacsat,manuscript=article,superscriptaddress,amsmath,amssymb,textcomp]{achemso}
\usepackage{chemformula}
\usepackage[T1]{fontenc}

\usepackage{graphicx}
\usepackage{dcolumn}
\usepackage{bm}

\usepackage{color,soul}
\usepackage{amsmath}
\usepackage{upgreek}
\usepackage{lipsum}  
\usepackage{comment}

\usepackage{hyperref}

\usepackage{xr}
\author{Jan Hidding}
\altaffiliation{These authors contributed equally}
\author{Cédric A. Cordero-Silis}
\affiliation{Zernike Institute for Advanced Materials, University of Groningen, 9747 AG Groningen, The Netherlands}
\altaffiliation{These authors contributed equally}

\author{Daniel Vaquero}
\affiliation{Nanotechnology Group, USAL—Nanolab, Universidad de Salamanca, E-37008 Salamanca, Spain}

\author{Konstantinos P. Rompotis}
\affiliation{Zernike Institute for Advanced Materials, University of Groningen, 9747 AG Groningen, The Netherlands}

\author{Jorge Quereda}
\affiliation{Departamento de Física de Materiales, GISC, Universidad Complutense de Madrid, E-28040 Madrid, Spain}

\author{Marcos H. D. Guimar\~{a}es}
\email{m.h.guimaraes@rug.nl}
\affiliation{Zernike Institute for Advanced Materials, University of Groningen, 9747 AG Groningen, The Netherlands}

\title{Locally Phase-Engineered MoTe\textsubscript{2} for Near-Infrared Photodetectors}


\begin{document}

\begin{abstract}
Transition metal dichalcogenides (TMDs) are ideal systems for two-dimensional (2D) optoelectronic applications, owing to their strong light-matter interaction and various band gap energies.
New techniques to modify the crystallographic phase of TMDs have recently been discovered, allowing the creation of lateral heterostructures and the design of all-2D circuitry.
Thus far, the potential benefits of phase-engineered TMD devices for optoelectronic applications are still largely unexplored. The dominant mechanisms involved in the photocurrent generation in these systems remain unclear, hindering further development of new all-2D optoelectronic devices.
Here, we fabricate locally phase-engineered MoTe\textsubscript{2} optoelectronic devices, creating a metal (1T’) semiconductor (2H) lateral junction and unveil the main mechanisms at play for photocurrent generation.
We find that the photocurrent originates from the 1T'\textendash2H junction, with a maximum at the 2H MoTe\textsubscript{2} side of the junction.
This observation, together with the non-linear IV-curve, indicates that the photovoltaic effect
plays a major role on the photon-to-charge current conversion in these systems.
Additionally, the 1T'\textendash2H MoTe\textsubscript{2} heterojunction device exhibits a fast optoelectronic response over a wavelength range of 700 nm to 1100 nm, with a rise and fall times of 113 \textmu s and 110 \textmu s, two orders of magnitude faster when compared to a directly contacted 2H MoTe\textsubscript{2} device.
These results show the potential of local phase-engineering for all-2D optoelectronic circuitry.
\end{abstract}

\maketitle
\section{Introduction}\label{sec:introduction}
The large family of two-dimensional (2D) materials called the transition metal dichalcogenides (TMDs) have gained much attention in the last two decades due to their versatility, mechanical strength, atomically flat interfaces, and strong absorption at the monolayer limit, making them promising candidates for future (opto)electronic and (opto)spintronic applications \cite{Buscema2015a}.
The most commonly studied crystal structure of the TMD family is the hexagonal (2H) phase, for which most TMDs are semiconducting, and possess a thickness-dependent band gap \cite{Mak2010, Manzeli2017}.
Apart from the 2H phase, however, TMDs can present a multitude of different crystallographic phases, such as the semiconducting 3R phase, or the semi-metallic 1T, 1T' and 1T\textsubscript{d} phases, which possess different symmetries and (opto)electronic properties.
To benefit from these different properties in a single device, researchers have recently focused on gaining control of the crystallographic phase of TMDs, allowing them to transform the phase of single TMD crystals at will.\cite{Aftab2023,Hu2023}
This new and emerging field is now referred to as the field of phase-engineering and opens the door to creating on-chip 2D circuitry with 2D metals and semiconductors.

In literature, multiple methods are used to induce a 2H to 1T' phase transformation for different TMDs, such as crystal deformation \cite{Dave2004, Song2016, Lin2014, Shang2019}, electrostatic doping \cite{Wang2017}, chemical doping \cite{Kappera2014a, Ma2015}, laser heating \cite{Cho2015}, etc.
In particular, MoTe\textsubscript{2} gained much attention as the energy barrier between the 2H and 1T' phase is the smallest ($\sim$40 meV) \cite{Duerloo2014}.
It was shown electrically that the Schottky barrier, present when directly contacting the 2H TMD with metallic contacts, is significantly reduced when contacting a 2H TMD via a phase transformed 1T' region \cite{Kappera2014a, Cho2015, Zhang2019, Bae2021}, with these two-dimensional lateral junctions, approaching the quantum limit for the contact resistance \cite{Wang2022}.
The possibility of fabricating high-quality contacts to a 2D semiconductor is essential to increase the speed of optoelectronic devices based on these materials.

Apart from electrical characterization, only a few reports explored the benefits of local phase engineering on the optoelectronic performance of TMD devices \cite{Yamaguchi2015, Lin2021, Ding2023}.
Lin et al. report an increased responsivity for 2H MoTe\textsubscript{2} devices using 1T' interlayer contacts \cite{Lin2021}.
However, no scanning photocurrent measurements are performed, which makes it difficult to disentangle the possible microscopic mechanism involved in the photocurrent generation to either the photovoltaic effect (PVE), due to the build in electric field at the Schottky barriers, or the photothermal effect (PTE), due to the different Seebeck coefficients of the 2H and 1T' region \cite{Xu2010, Zhang2015, Buscema2015a}.

Here, we perform scanning photocurrent measurements on 1T'\textendash2H MoTe\textsubscript{2} heterojunction devices, which allow us to spatially resolve the areas involved in the photocurrent generation, giving insights on the underlying mechanisms involved.
First, we phase-transform the sides of an exfoliated 2H MoTe\textsubscript{2} crystal to a 1T' phase using local heating by laser irradiation, which allows us to contact the 2H MoTe\textsubscript{2} via the semi-metallic 1T' regions.
We find a clear non-linear behavior for the 1T'\textendash contacted 2H region, indicative of a Schottky barrier between the 1T' and 2H region.
Additionally, using the scanning photocurrent measurements, we clearly observe that the photocurrents are generated at the 1T'\textendash2H junction rather than at the Ti/Au electrodes or the 1T' region.
More specifically, we find that the peak of the photocurrent is generated at the 2H side of the junction, which suggests that the observed photocurrents in the 1T'\textendash2H junction can be attributed to the PVE rather than the PTE.
Lastly, we characterize the optoelectronic performance of the MoTe\textsubscript{2} photodetector by performing time-resolved, and laser power dependent photocurrent measurements.
We find fast rise and fall times of 113 \textmu s and 110 \textmu s, respectively, over a broad spectral range of 700 nm to 1100 nm.
By comparing our 1T'\textendash2H MoTe\textsubscript{2} photodetector to a 2H MoTe\textsubscript{2} diode where the electrodes are directly deposited on the 2H MoTe\textsubscript{2} crystal, we are able to show that the temporal response of 1T'\textendash contacted 2H MoTe\textsubscript{2} is two orders of magnitude faster.
This indicates that phase-engineering can be considered another tool for improving the performance of TMD-based optoelectronic devices.

\section{Results and Discussion}\label{sec:results}
\begin{figure*}[t!]
	\includegraphics{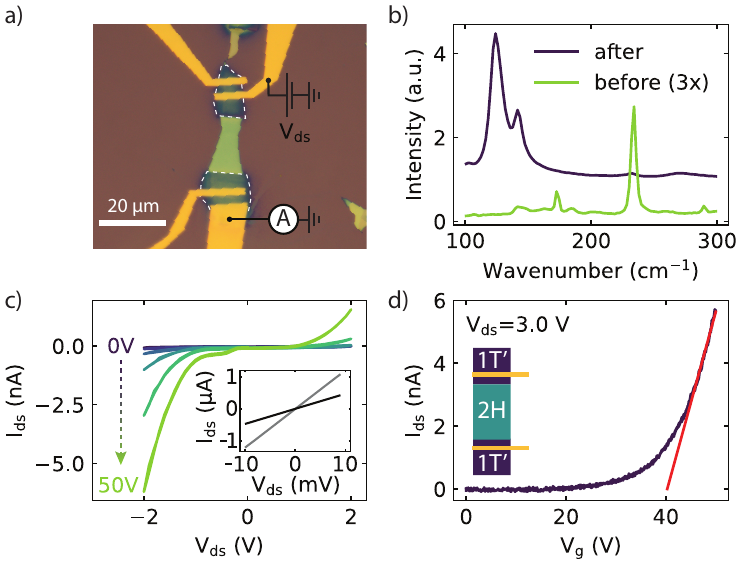}
	\caption{(a) Optical micrograph of a phase-changed MoTe\textsubscript{2} device, where the phase changed regions are outlined with the white dashed line while the bright green part is the unaltered 2H MoTe\textsubscript{2} region. (b) The Raman spectra obtained before (green) and after (purple) the phase transformation, which clearly indicate a successful phase transformation. The spectra before the phase change is multiplied by 3 for clarity. (c) The $I_{ds}-V_{ds}$ measurements, as indicated in (a), with $V_{g}$ ranging from 0 V to 50 V, taken at 78 K. The non-linear IV characteristics show the Schottky behavior. The IV measurement for the two 1T' regions are depicted in the inset, which clearly show Ohmic behavior. (d) The transfer curve measured with a $V_{ds}$ of 3 V, taken at 78 K, shows clear n-type behavior.}
	\label{fig:figure1}
\end{figure*}

\subsection{Raman Spectroscopy}
The device used to perform the optoelectronic measurements is depicted in Fig. \ref{fig:figure1}(a).
The green region is the untreated 2H MoTe\textsubscript{2} while the dark green regions, indicated by the dashed white line, were irradiated with a laser to induce the phase transformation from 2H to 1T' (details of the device fabrication can be found in the Methods section).
To confirm the phase transformation, we performed Raman spectroscopy measurements as depicted in Fig. \ref{fig:figure1}(b).
Before laser irradiation, we observe the in-plane $E_{2g}$ mode at 235 cm$^{-1}$ and an out-of-plane $A_{g}$ mode near 174 cm$^{-1}$, indicative of the 2H MoTe\textsubscript{2} phase.
After laser irradiation, these peaks disappear, and we observe two new peaks at 124 cm$^{-1}$ and 138 cm$^{-1}$, corresponding to the $A_{g}$ mode of 1T' MoTe\textsubscript{2}.
This significant change in the Raman spectrum indicates the successful phase transformation of the irradiated regions.\cite{Cho2015,Tan2018}

\subsection{Electrical Characterization}
After fabricating electrical contacts to the phase-changed region, we electrically characterize the device in a cryostat at 78 K. We swept the drain-source voltage ($V_{ds}$) and measured the drain-source current ($I_{ds}$) for the different regions (both 1T' and the 1T'\textendash2H junction).
In Fig. \ref{fig:figure1}(c), the 2-probe $I_{ds}$-$V_{ds}$ measurements are shown for the 1T'\textendash2H\textendash1T' junction at different gate voltages, ranging from 0 V to 50 V.
We observe a clear non-linear behavior for the $I_{ds}$ as function of the $V_{ds}$, indicative of a Schottky barrier present in our device, which could either be between the Ti/Au contact and the 1T'\textendash MoTe\textsubscript{2}, or the 1T'\textendash2H junction.
To determine this, we performed the $I_{ds}$-$V_{ds}$ measurement on the phase transformed 1T' region only and observe a clear linear behavior showing Ohmic contact between the Ti/Au contacts and 1T' region, depicted in the inset of Fig. \ref{fig:figure1}(c).
Therefore, we expect the non-linear behavior observed in Fig. \ref{fig:figure1}(c) to originate from a Schottky barrier between the 1T'\textendash2H junction.
For the 1T' regions, we find a 2-probe resistance of 8 k$\Omega$ and 18 k$\Omega$, which again indicate the successful transformation from the semiconducting 2H MoTe\textsubscript{2} phase to the semi-metallic 1T' phase.

\begin{figure*}[t]
	\centering
	\includegraphics{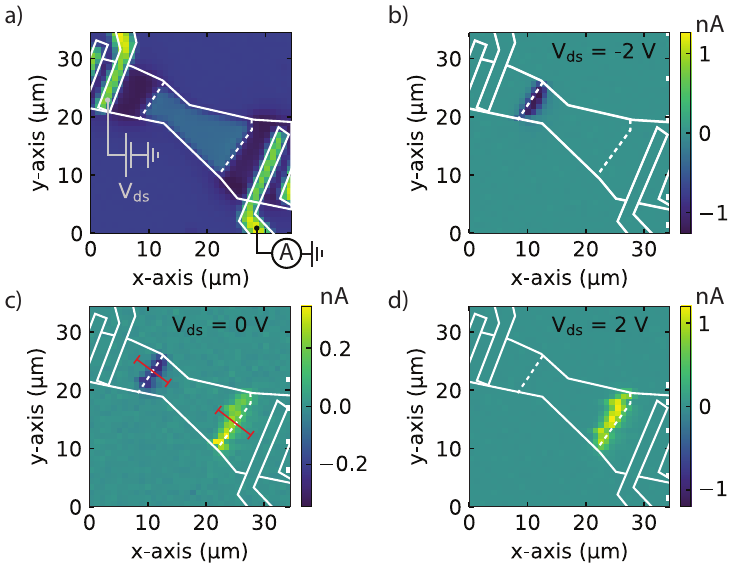}
	\caption{(a) Reflectivity map of the scanning photocurrent measurement of the device depicted in Fig. \ref{fig:figure1}(a) with the corresponding photocurrent map in (b), (c), and (d), taken at RT. The white outlines indicate the position of the flake and Ti/Au contacts, while the white dashed lines indicates the 1T'\textendash2H junctions for clarity. The photocurrent maps are obtained with $\lambda=700$ nm, $P=1$ \textmu W, and a $V_{ds}$ of (b) -2 V, (c) 0 V, and (d) 2 V. From the photocurrent maps, we can clearly see that the induced photocurrent originates from the 1T'\textendash2H junction rather than from the Ti/Au contacts.}
	\label{fig:figure2}
\end{figure*}

\subsection{Optoelectrical Characterization}
To observe the optoelectronic response of the 1T'\textendash2H\textendash1T' sample, we perform both scanning photocurrent measurements and time-resolved photocurrent measurements, as described in the methods section.
First, the scanning photocurrent measurements are presented, which give more insight on the origin of the photocurrent, after which the time-resolved photocurrent measurements are discussed.

The scanning photocurrent measurements enable us to spatially identify where the photocurrent is generated, and thus allows us to check whether the photocurrent originates from the Ti/Au contacts or from the MoTe\textsubscript{2} flake itself.
When performing the scanning photocurrent measurements, the laser beam is focused and scanned across the sample in a raster-like fashion while recording both the reflection and the generated photocurrent.
Figure \ref{fig:figure2}(a) shows the recorded reflection map with an illumination wavelength of 700 nm, a power of 1 \textmu W, and a full width half maximum (FHWM) spot size of $0.70\pm0.02$ \textmu m (see supporting information).
The contours of the flake and the Ti/Au contacts are clearly visible and highlighted with the white outlines for clarity.
Note that the different phases of the MoTe\textsubscript{2} can also be clearly distinguished, and their junctions are highlighted with the white dashed line.
The corresponding photocurrent map with a $V_{ds}$ of -2 V, 0 V, and 2 V are depicted in Fig. \ref{fig:figure2}(b), (c), and (d), respectively.
We clearly observe that the photocurrent originates locally from the 1T'\textendash2H junction, rather than the Ti/Au contacts, which is the case when directly contacting the 2H MoTe\textsubscript{2}.
This, again, confirms a low Schottky barrier between the Ti/Au and 1T' MoTe\textsubscript{2}, indicating a successful phase transformation.

\begin{figure}[t]
	\includegraphics{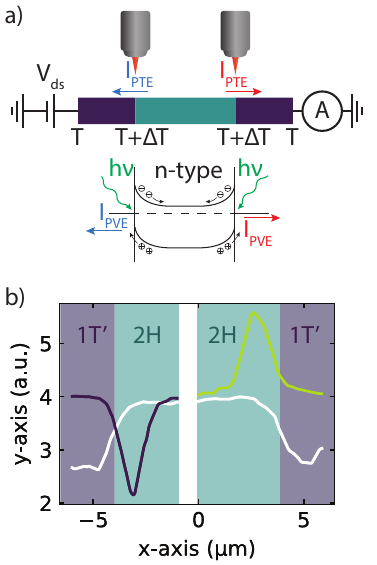}
	\caption{(a) A schematic of the photo-thermoelectric (PTE) and photovoltaic effect (PVE). For the PTE, the laser locally heats the device which creates a temperature gradient, which via the Seebeck effect causes an induced photocurrent ($I_{PTE}$). For the PVE, the localized electric field at the Schottky barrier causes a separation of the photo-induced carriers, resulting in $I_{PVE}$. Note that the two effect produce a photocurrent with the same direction, and that the induced photocurrent is opposite on both junctions. (b) A line trace of the photocurrent mapping, indicated in black in Fig. \ref{fig:figure2}(c), showing the reflection (white) and negative (purple) and positive (green) photocurrent peaks along the line trace. Both the maximum and minimum photocurrent are obtained in the 2H region, as expected for a localized electric field from a Schottky barrier. The phases are indicated by the purple (1T') and green (2H) background.}
	\label{fig:figure3}
\end{figure}

\begin{figure*}[t!]
	\includegraphics{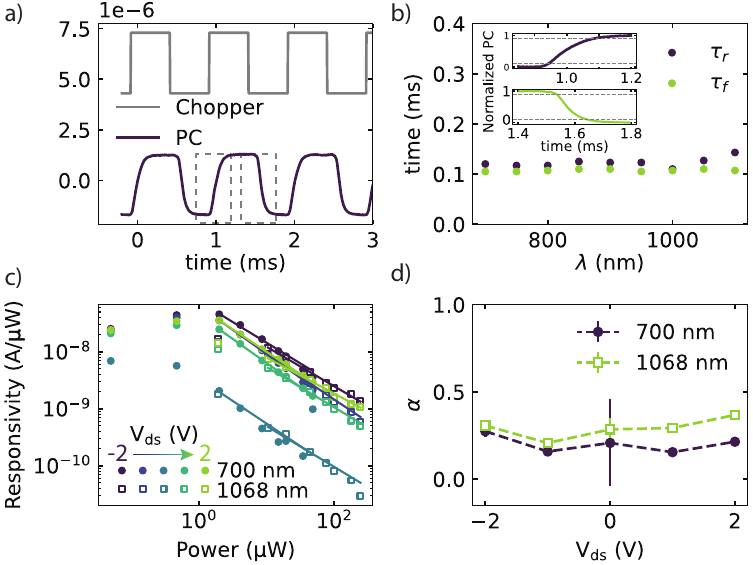}
	\caption{(a) Time-resolved photocurrent, taken at RT, where the photocurrent (purple) in the device is plotted together with the chopper signal (grey) versus time show the fast response of our MoTe\textsubscript{2} photodetector. The dashed grey rectangles indicate the region used to determine the rise and fall time, as depicted in the inset of (b). (b) The extracted rise (purple) and fall times (green) indicate no wavelength dependence on the fast response for wavelengths ranging from 700 to 1100 nm. The inset shows the rise (purple) and fall (green) curves of the photocurrent from which the rise and fall times are extracted. (c) The power dependent measurements for different $V_{ds}$, ranging from -2 V to 2 V, with a maximum responsivity of $4.5\times 10^{-8}$ A/\textmu W. Here, the responsivity ($R$) of the device is plotted as function of the laser excitation power ($P$) and fitted at high laser excitation power to a power law: $R\propto P^{\alpha-1}$. The measured $R$ for 700 nm and 1068 nm are indicated by the filled circles and unfilled squares, respectively. (d) The extracted index of the power law ($\alpha$) from the fitting in (c) versus $V_{ds}$ for the wavelengths 700 nm (purple) and 1068 nm (green).}
	\label{fig:figure4}
\end{figure*}

We observe photocurrents with opposite signs at the two 1T’-2H junctions for $V_{ds}$ = 0 V.
This is in line with the expected behavior for two possible mechanisms: photovoltaic effect (PVE) due to a Schottky barrier at the 1T'\textendash2H junction, or the photo-thermoelectric effect (PTE) due to a different Seebeck coefficient of the two MoTe\textsubscript{2} phases.
The two mechanisms are schematically depicted in Fig. \ref{fig:figure3}. While both the PVE and the PTE can give rise to photocurrent in the 2H-1T' interface, we conclude that the PTE contribution should be negligible. For the PTE, local heating due to the laser irradiation causes a temperature gradient ($\Delta T$) which is converted into a voltage difference ($V_{PTE}$) due to a difference in the Seebeck coefficient between the 1T' ($S_{\mathrm{1T'}}$) and 2H ($S_{\mathrm{2H}}$) phase \cite{Buscema2015a, Huo2018}. By using the maximum generated photocurrent in our measurements, we calculate a range of unrealistic temperature gradients above $7000$ K, indicating that the PTE cannot, solely, explain the observed photocurrent. A more detailed explanation can be found in the supporting information.

On the other hand, for the PVE driven photocurrent, the localized electric field at the 1T'\textendash2H interface causes the photo-induced electron-hole pairs to separate, resulting in a photocurrent \cite{Huo2018}.
For our n-type MoTe\textsubscript{2}, the band alignment is depicted in Fig. \ref{fig:figure3}(a).
The electric field from the Schottky barrier is positioned in the 2H-region.
By taking a line scan of the reflection map and photocurrent map, indicated by the red line in Fig. \ref{fig:figure2}(c), we can more accurately determine the position of the photocurrent peak with respect to the 1T'\textendash2H junction.
Here, we find that the peak of the photocurrent arises in the 2H region rather than at the 1T'\textendash2H junction, which is in line with the expectation for the PVE \cite{Zhang2015}.
Additionally, we can estimate the depletion region due to the Schottky barrier by using the following equation:

\begin{equation}
    \label{eq:depletionwidth}
    W=\sqrt{\frac{2\epsilon_{0}\epsilon_{r}\phi_{bi}}{e N_{d}}},
\end{equation}

\noindent where $W$ is the width of the Schottky barrier, $e$ is the electron charge, $\epsilon_{0}$ is the permittivity of free space.
By assuming a donor density $N_{d}$ of $10^{11}$ cm$^2$ \cite{HQgraphene}, a barrier height $\phi_{bi}$ of 60 meV \cite{Lin2021} and a relative $\epsilon_{r}$ of 12, we estimate $W$ to be $\sim2$ \textmu m, which corresponds well to the FWHM of $1.5\pm0.2$ \textmu m we find by fitting the positive photocurrent peak with a Gaussian.

Buscema et al. observe a similar photocurrent sign in their scanning photocurrent measurements on an n-type MoS\textsubscript{2} photodiode \cite{Buscema2013}.
However, they attribute the induced photocurrent to the PTE, as they observe a clear photocurrent generation in the center of their Ti/Au contacts and see a linear $I_{ds}$-$V_{ds}$ behavior with no indication of a Schottky barrier.
In contrast, for our 1T’-2H MoTe2 junctions, we observe a clear non-linear IV-curve, indicating that the Schottky barrier plays a more important role in our devices.
Furthermore, they see a pronounced photocurrent even when exciting below the bandgap of MoS\textsubscript{2}.
Unfortunately, our optoelectronic setup only allows for excitation up to 1100 nm, which is still above the bandgap of MoTe\textsubscript{2} ($\sim$1.1 eV $\propto$ $\sim$1127 nm for bulk) \cite{Ruppert2014, Lezama2015}.
Therefore, we suggest further research to be performed on below band gap excitation to determine to what extent the PTE is contributing to the observed photocurrent.

To characterize the optoelectronic performance of our MoTe\textsubscript{2} photodetector, we perform time-resolved and power-dependent photocurrent measurements, depicted in Fig. \ref{fig:figure4}.
By measuring the induced photocurrent versus time, using a chopper to modulate the light on and off (see Fig. \ref{fig:figure4}(a)), we are able to extract the rise ($\tau_{r}$) and fall time ($\tau_{f}$)) of the device, which are defined as the time required for the photocurrent to increase from 10\% to 90\%, and decrease from 90\% to 10\% of the maximum photocurrent, respectively.
By performing these measurements over a range of different excitation wavelengths, we find that we get a short rise and fall time of $\sim$113 \textmu s and $\sim$110 \textmu s, respectively, independent of the wavelength as depicted in Fig. \ref{fig:figure4}(b).
These response times correspond to a 3 dB frequency of $0.35/\tau_{r} = 3$ kHz \cite{PinkiYadav2022}, which are close to the performance of graphene/MoTe\textsubscript{2}/graphene photodetectors \cite{Zhang2017}.
In contrast, when directly contacting the 2H MoTe\textsubscript{2} with Ti/Au electrodes, we find a much slower response, as shown in Fig. S2(c) in the supporting information.
Here, we observe a waveform similar to a ``\textit{sawtooth}'' response, indicative of capacitive behavior, already at 20 Hz. This is a result of the highly resistive contact with the 2H region of the device, leading to photogating and slow charging of the device.
This shows that using the 1T' regions to contact the 2H MoTe\textsubscript{2} increases the temporal response of our MoTe\textsubscript{2} photodetector by more than two orders of magnitude.

The response dynamics displayed at the 1T'\textendash2H junction are fast compared to other TMD-based photodetectors \cite{Xiao2022, PinkiYadav2022}.
More specifically, compared to other MoTe\textsubscript{2} based photodetectors, they are one order of magnitude faster than the report of Huang et al. on 2H MoTe\textsubscript{2} \cite{Huang2016}, and similar to the ones found by Lin et al. in 1T'\textendash contacted 2H MoTe\textsubscript{2} \cite{Lin2021}.
On other TMD-based devices, a variety of different response dynamics are reported, with the fastest responses reported for deep UV and mid-IR detectors on graphene/MoTe\textsubscript{2}/black phosphorus devices, which reach bandwidths of 2.1 MHz \cite{Shen2022}.

To determine the responsivity of our devices, we vary the excitation power at a fixed excitation wavelength (700 nm and 1068 nm).
From these power dependent measurements, we are able to determine the responsivity by: $R=I_{PC}/P$, where  $I_{PC}$ is the induced photocurrent and $P$ is the power of the laser \cite{Buscema2015a, Huo2018},  and find a maximum $R$ of $4.5\times 10^{-8}$ A/\textmu W with a wavelength of 700 nm and a 2 V bias.
The value we find is comparable to other reports on TMD-based photodetectors, ranging approximately from $7.25\times 10^{-11}$ A/\textmu W to $9.708\times 10^{-3}$ A/\textmu W \cite{Xiao2022,Yu2017,Huang2016,Yin2016}.
Given that there is commonly a trade-off between fast response and high responsivity in these devices \cite{PinkiYadav2022}, it is not unexpected that the large responsivity of our devices falls within the middle or lower half of the reported range.
Additionally, the responsivity of our device is measured with a focused laser spot rather than illuminating the entire device.
This could lead to an underestimation of the generated photocurrent as only a small fraction of the photodetector area is used to generate the photocurrent.

Figure \ref{fig:figure4}(c) clearly shows a decrease of responsivity with incident excitation power for $P>1.9$ \textmu W, which is commonly observed in TMD photodetectors \cite{Lee2012, Lopez-Sanchez2013}.
It can be associated with a reduced number of photogenerated carriers available for extraction under high photon flux due to the saturation of recombination/trap states that influence the lifetime of the generated carriers \cite{Buscema2014}.
The responsivity versus laser power can be expressed by a power law $R\propto P^{\alpha-1}_{d}$ for $P>1.9$ \textmu W, where $P$ is the laser power, and $\alpha$ is the index of the power law \cite{Huang2016, Zhang2016d}.
From the fit, we are able to extract $\alpha$ for the two different wavelengths at different $V_{ds}$ as shown in Fig. \ref{fig:figure4}(d).
The deviation from the ideal slope of $\alpha = 1$, where the responsivity does not dependent on the laser power, can be attributed to complex processes in the carrier generation, trapping, and electron-hole recombination in the MoTe\textsubscript{2} \cite{Kind2002, Liu2014}.
For the PTE, a value of $\alpha\sim0.8$ is expected, while we find a value of $\alpha\sim0.25$, which indicates again that the PTE is not primarily responsible for the generated photocurrent in our device \cite{Xu2010}.

\section{Conclusion}\label{sec:conclusion}
In conclusion, our results indicate that contacting the 2H region of MoTe\textsubscript{2} via a phase-transformed 1T' region is beneficial for the temporal optoelectronic response of MoTe\textsubscript{2} based photodetectors.
Our scanning photocurrent measurements and non-linear IV curves, clearly show that the origin of the photocurrent in our devices can be ascribed to the Schottky barrier between the 1T'\textendash2H junction, rather than the photo-thermoelectric effect or Schottky barriers at the Ti/Au electrode-TMD interface.
Contacting the MoTe\textsubscript{2} via the phase-transformed 1T' region, therefore, allows one to study the intrinsic properties of the TMD rather than the electrode-TMD interactions, beneficial for fundamental research.
Additionally, an increase of 2 orders of magnitude in the optoelectronic temporal response is observed when contacting the 2H MoTe\textsubscript{2} via the 1T' regions.
This shows that tailoring the crystallographic phase of TMDs locally, altering their optoelectronic response at will, can have a profitable effect on the optoelectronic operation.
Our results, in combination with the wide variety of phase-engineering techniques and different TMDs available, could lead to a further improved performance of TMD-based optoelectronic devices, leading to more sensitive, faster and flexible photodetectors.

\section{Methods}\label{sec:methods}
\subsection{Device fabrication}
The 2H MoTe\textsubscript{2} flakes are obtained by mechanical exfoliation (bulk crystal supplied by HQ Graphene) and transferred onto a Si/SiO\textsubscript{2} (285 nm) substrate in a nitrogen environment.
Using an optical microscope, the MoTe\textsubscript{2} flakes are selected based on their size, thickness, and homogeneous surface.
Next, the Raman spectra are obtained with an inVia Raman Renishaw microscope using a linearly polarized laser in backscattering geometry. The excitation wavelength and grating used were $\lambda=532$ nm and $2400$ l/mm respectively. The laser power was $\sim 100$ $\upmu \mathrm{W}$ with a diffraction-limited spot of $\sim 1$ \textmu m.
Using the same system, the 2H-1T' phase transformation is performed by selectively illuminating parts of the MoTe\textsubscript{2} flake with the 532 nm laser beam in a raster-like fashion, using steps of 500 nm and 0.1 s illumination.
We find that a laser power $\geq3.25$ $\mathrm{mW}$ (laser spot size around 500 nm) is needed to initiate the phase transformation.
Finally, using standard lithography techniques, the Ti/Au (5 nm/55 nm) contacts are fabricated on top of the flake by means of electron beam lithography and electron beam evaporation.

\subsection{Optoelectronic measurements}
The electrical characterization (i.e. IV-sweeps, transfer curves) is performed using Keithleys 2400 and 2450 source measure units at 78 K.
For the optoelectronic measurements, a supercontinuum white light laser (NKT Photonics SuperK EXTREME) is used as illumination source, and the measurements are taken at room temperature.
The induced photocurrent is measured in a short circuit configuration using a Stanford Research Systems SR830 lock-in amplifier, which is referenced to the frequency of the optical chopper.
The photocurrents are either measured directly by the lock-in amplifier, or converted to a voltage using a home build current pre-amplifier, which is subsequently measured by the lock-in amplifier.
The time-resolved photoresponse of the device, depicted in Fig. \ref{fig:figure3}(a), is measured using a chopper and an oscilloscope (Keysight DSOX1204A) at room temperature. 

\section{Acknowledgements}
The authors acknowledge Prof. M. A. Loi and E. K. Tekelenburg for their help with the Raman measurements and thank J. G. Holstein, H. Adema, H. de Vries, A. Joshua, and F. H. van der Velde for their technical support.
Sample fabrication was performed using NanoLabNL facilities.
This work was supported by the Dutch Research Council (NWO—STU.019.014), the European Union (ERC, 2D-OPTOSPIN, 101076932), the ``Materials for the Quantum Age'' (QuMat) program (Registration No. 024.005.006) which is part of the Gravitation program financed by the Dutch Ministry of Education, Culture and Science (OCW), the Zernike Institute for Advanced Materials, and the innovation program under grant agreement No. 881603 (Graphene Flagship).

\section{Author contributions}
J.H. and C.A.C.S. fabricated the samples, and together with D.V. performed both the electrical and optical measurements under the supervision of M.H.D.G.. K.R. joined to perform measurements and sample characterization on additional samples also under the supervision of M.H.D.G..
J.H. performed the data analysis and, together with C.A.C.S and M.H.D.G., wrote the paper with comments from all authors.

\section{Supporting Information}
The Supporting Information contains: details over the Photothermoelectric temperature gradient calculation; the laser spot size calculation and data for 700 and 1064 nm for the reported measurements; equation and additional details for charge carrier mobility and further details of another phase-engineered device, including responsivity and transfer curves.

\clearpage

\bibliography{bibliography}


\clearpage


\end{document}


\clearpage


\section{PTE temperature gradient calculation}

In a short-circuit configuration, laser irradiation causes a temperature gradient that can lead to an induced photocurrent driven by the PTE. By using the Seebeck coefficient of the 2H and 1T' phases which were shown to be, respectively, $S_{2H}\sim$ 230 \textmu V K$^{-1}$ and $S_{1T'}\sim$ 30 \textmu V K$^{-1}$ \cite{Keum2015}, we can calculate an approximate temperature gradient for our devices using the following equation:
\begin{equation}
    V_{PTE}=(S_{\mathrm{1T'}}-S_{\mathrm{2H}}) \Delta T.
    \label{eq:Vpte}
\end{equation}

Assuming that the maximum photocurrent solely originates from the PTE, the local temperature gradient can be calculated by considering the photocurrent generated at the 2H-1T' interfaces in Fig. 2(c), and the 2-probe resistance through the channel. Together with the values for the Seebeck coefficient mentioned above, we find unrealistically high temperature gradients of between $7490$ K and $11952$ K. This is indicative that the PTE is not the sole mechanism responsible for the generated photocurrent.

\clearpage

\section{Laser spot determination}

\begin{figure*}[ht]
    \makeatletter 
    \renewcommand{\thefigure}{S\@arabic\c@figure}
    \makeatother
    \setcounter{figure}{0}
	\includegraphics{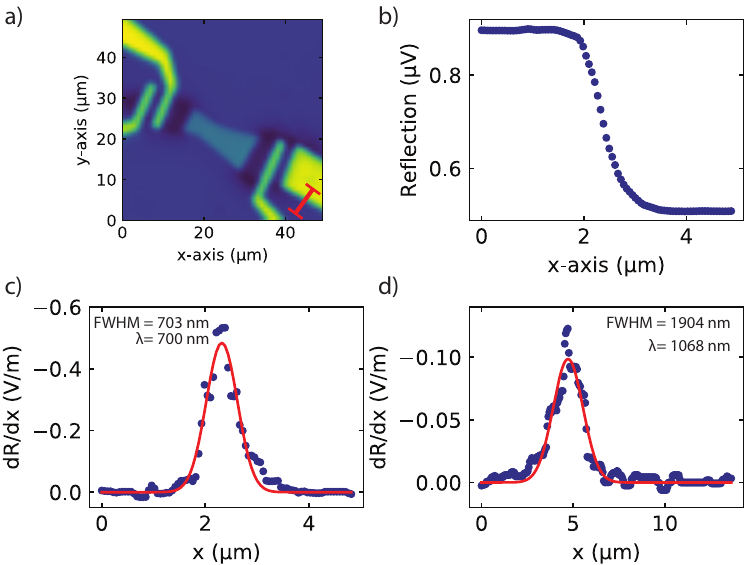}
	\caption{(a) Reflection map taken with $\lambda=700$ nm of the device depicted in Fig. 1(a), where the red line indicated the position of the line trace to determine the laser spot size. (b) The data (blue circles) of the line trace in (a), which shows the clear step from the Ti/Au electrode to the SiO\textsubscript{2} substrate. (c) By taking the derivative of the data depicted in (b) (blue circles), and fitting it with a Gaussian (red line), the FWHM of the laser spot used to do the scanning is determined. (d) Same as (c) but now for $\lambda=1068$ nm}.
	\label{fig:figure5}
\end{figure*}

To determine the spot size of our laser at the wavelengths of 700 nm and 1068 nm, we use the software Gwyddion to take a line trace over a scanning map of the reflectivity.
The reflectivity map for 700 nm is depicted in Fig. \ref{fig:figure5}(a), where the position of the line trace is indicated with a red line and the line trace itself is depicted in Fig. \ref{fig:figure5}(b).
By fitting the derivative of the data in Fig. \ref{fig:figure5}(b) with a Gaussian, we are able to determine the FWHM of the laser spot, as show in Fig. \ref{fig:figure5}(c) and (d) for 700 nm and 1068 nm, respectively.
For a wavelength of 700 nm, we find a diffraction limited FWHM of $0.70 \pm 0.02$ \textmu m, while for the 1068 nm, we find a broader FWHM of $1.91 \pm 0.04$ \textmu m.

\clearpage

\section{Mobility}

The transfer curve of Fig. 1(d), for the device depicted in Fig. 1(a), shows clear n-type behavior with a threshold voltage of $V_{th}=40.3$ V.
The mobility $\mu$ of the device is determined by:

\begin{equation}
    \label{eq:mobility}
    \mu=\left(\frac{dI_{ds}}{dV_{g}}\right) \left(\frac{l}{wC_{g}V_{ds}}\right),
\end{equation}

\noindent where $\left(dI_{ds}/V_{g}\right)$ is the slope at positive $V_{g}$, $l$ and $w$ are the length and width of the channel, $C_{g}$ is the area capacitance of the SiO\textsubscript{2} back gate ($1.2\times10^{-4}$ F/m$^{2}$), and $V_{ds}$ is the drain-source voltage (3 V).
We calculate a mobility of
$0.08$ $\mathrm{cm^{2}/(V\cdot s)}$.

\clearpage

\section{Other Phase-Engineered MoTe\textsubscript{2} Devices}
We measured other phase transformed MoTe\textsubscript{2} devices, one of them exhibits clear p-type behavior, and the other one n-type. For the p-type device, different mobilities compared to the device depicted in Fig. 1(a) are obtained. The difference is attributed to the following explanation: When contacting the 2H MoTe\textsubscript{2} directly with the Ti/Au contacts, we find a mobility of $1.74$ $\mathrm{cm^2/V\cdot s}$, while if the 2H region is contacted via the 1T' region, similar to Fig. 1(d), we find a mobility of $14.18$ $\mathrm{cm^2/V\cdot s}$, in agreement with the reports of Bae et al. \cite{Bae2021} on electrical measurements on phase changed MoTe\textsubscript{2} devices and larger compared to similar measurements reported  by Zhang et al. \cite{Zhang2019}.

Finally, the n-type phase changed MoTe\textsubscript{2} device, with the same contact geometry as the device in Figure 1(d), displays a relatively high photoresponse that, despite being one order of magnitude lower to the device in Figure 1(d), is still comparable to previously reported devices \cite{Huang2016}. As in the case for Figure 4(a), the calculated rise and fall times of this device are, respectively, $\tau_r=1.26$ ms and $\tau_f= 1.12$ ms, with a 3 dB frequency of 0.277kHz.

\begin{figure*}[t]
    \makeatletter 
    \renewcommand{\thefigure}{S\@arabic\c@figure}
    \makeatother
    \includegraphics[width=0.7969\textwidth]{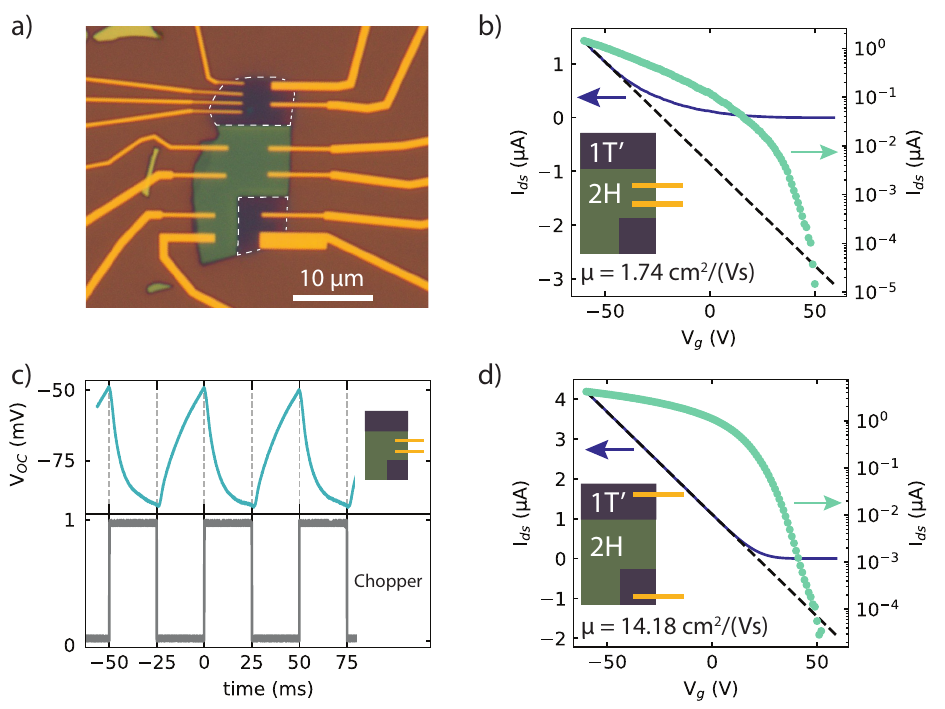}
	\caption{(a) Optical micrograph of another phase-engineered MoTe\textsubscript{2} device similar to the device depicted in Fig. 1(a) in the main text. The white dashed regions indicate the areas where the 2H MoTe\textsubscript{2} are transformed to 1T' by laser irradiation. (b) Transfer curve measured with the Ti/Au electrodes directly deposited on the 2H region, which shows clear p-type behavior. (d) The transfer curve measured with the Ti/Au electrodes on the 1T' region. By fitting the curve on negative gate voltages and using Eq. \ref{eq:mobility}, we are able to extract a mobility of 1.74 $\mathrm{cm^{2}/(V\cdot s)}$ and 14.2 $\mathrm{cm^{2}/(V\cdot s)}$, respectively. (c) The top panel shows the temporal photovoltage response (green) of the MoTe\textsubscript{2} device when the 2H crystal is directly contacted with the Ti/Au electrodes. The bottom panel depicts the signal from the chopper (grey), which chops the light on and off. A much slower optoelectronic response is observed compared to the response of the device discussed in the main text, as shown in Fig. 4(a).}
	\label{fig:figure6}
\end{figure*}

\clearpage



\bibliography{bibliography}